\documentclass{article}
\usepackage[margin=20mm]{geometry}
\usepackage[margin=20mm]{caption}
\usepackage{hyperref}

\usepackage{graphicx}     
\usepackage{ragged2e}

\usepackage[bottom]{footmisc} 
\usepackage{amsmath,amssymb,amsfonts}        
\usepackage{nicematrix}
 \usepackage{mathtools}
\usepackage[noadjust]{cite}
\usepackage[utf8]{inputenc}
\usepackage[T1]{fontenc}

 \usepackage{bm}

\usepackage{tikz,capt-of}
\usetikzlibrary{arrows,matrix,positioning,shapes,decorations.pathreplacing,shapes.misc}
\usetikzlibrary{fit}
\usetikzlibrary{calc}
\usetikzlibrary{arrows,matrix,positioning,shapes,patterns}
\usetikzlibrary{decorations.pathmorphing,backgrounds,decorations.pathreplacing} 

 \makeindex

 \newcommand{\bth}{\bm{\theta}}
 \newcommand{\bph}{\bm{\phi}}
\newcommand{\D}{\mathrm{d}}
\newcommand{\E}{\mathrm{e}}
\renewcommand{\vec}[1]{\mathbf{#1}}

\usepackage{amsthm}
\newtheorem{convention}{Convention}
\newtheorem{property}{Property}
\newtheorem{definition}{Definition}
\newtheorem{remark}{Remark}
\newtheorem{proposition}{Proposition}

\title{Model structures and structural identifiability: What? Why? How?\footnote{This is a pre-publication version of a paper to be published in the 2019-20 MATRIX Annals, Springer.}}
\author{Jason M. Whyte \\
ACEMS, School of Mathematics and Statistics, and  CEBRA, School of BioSciences,  \\ University of Melbourne, Parkville, Victoria, Australia, 3010}
\date{30th December 2019}

\begin{document}
\maketitle

\abstract{We may attempt to encapsulate what we know about a physical system by a model structure, $S$.
This collection of related models is defined by parametric relationships between system features; say observables (outputs), unobservable variables (states), and applied inputs. Each parameter vector in some parameter space is associated with a completely specified model in $S$.
Before choosing a model in $S$ to predict system behaviour, we must estimate its parameters from system observations.
Inconveniently, multiple models (associated with distinct parameter estimates) may approximate data equally well. Yet, if these equally valid alternatives produce dissimilar predictions of unobserved quantities, then we cannot confidently make predictions. Thus, our study may not yield any useful result. \newline\indent
We may anticipate the non-uniqueness of parameter estimates ahead of data collection by testing $S$ for 
structural global identifiability (SGI). Here we will provide an overview of  the importance of SGI, some essential theory and distinctions, and demonstrate these in testing some examples.}

\section{Introduction} \label{sec:1}
A ``model structure'' (or simply ``structure'') is essentially a collection of related models of some particular class (say the linear, first-order, homogeneous, constant-coefficient ODEs in $n$ variables), as summarised by
mathematical relationships between system variables that depend on parameters. 
For example, in a ``controlled state-space structure'' we may draw on our knowledge of the system to relate time-varying quantities such as ``states'' (${\bf x}$) that we may not be able to observe, and (typically known) controls or ``inputs'' (${\bf u}$) which act on some part of our system, to ``outputs'' (${\bf y}$) we can observe. 
A structure is a useful construct when seeking to model some physical system 
for which our knowledge is incomplete. We choose some suitable parameter space, and each parameter vector therein is
associated with a model in our structure, where we use ``model'' to mean a completely specified set of mathematical relationships between system variables. 

In order to illustrate the concept of a structure, we will consider $S_{1}$, a controlled state-space structure of ``compartmental'' models, meaning that these are subject to a ``conservation of mass'' condition---matter is neither created nor destroyed. 
When we are interested in a system evolving in continuous time, a structure will employ ordinary differential equations (ODEs) to
describe the time course of the states. Compartmental structures are often appropriate for the modelling of biological systems. 
To illustrate this, let us consider a simple biochemical system, where we consider the interconversion and consumption of chemical species, as in a cellular process.
Structure $S_{1}$ has three states, $x_{1}$, $x_{2}$, and $x_{3}$, representing concentrations of three distinct chemical species, or ``compartments''. Matter may be excreted from the system, delivered into the system,  or converted between the forms. 
We assume that the system receives some infusion of $x_{3}$ via input $u$.

Using standard notation for compartmental systems, a real parameter $k_{ij}$ ($i,j = 1,2,3$, $i \neq j$) represents the rate constant for the conversion of $x_{j}$ into $x_{i}$. A real parameter $k_{0j}$ is the rate constant associated with the loss of material from $x_{j}$ to the ``environment'' outside of the system.  If reactions are governed by ``first-order mass-action kinetics'', the rate of conversion (or excretion) 
of some species at time $t$ depends linearly on the amount of that species at time $t$.

Given our physical system and modelling paradigm, (and understanding that an expression such as $\dot{x}$  represents $\D x/\D t$) we may write the ``representative model'' of $S_{1}$ as
\begin{gather}
\begin{aligned}
\dot{x}_{1}(t) & = - (k_{01} + k_{21}) x_{1}(t) - k_{12}x_{2}(t) \; , \\
\dot{x}_{2}(t) & =  k_{21} x_{1}(t) - (k_{12} + k_{32}) x_{2}(t) + k_{23}x_{3}(t) \; , \\
\dot{x}_{3}(t) & =  k_{32} x_{2}(t) - k_{23}x_{3}(t) + u(t) \;  , 
\end{aligned} \label{eq:s1_example}\\
\intertext{where we set initial conditions for our states (where $\top$ denotes transpose)}
\begin{pmatrix} x_{1}(0) & x_{2}(0) & x_{3}(0) \end{pmatrix}^{\top}  = \begin{pmatrix} 0 & x_{2_{0}} & 0 \end{pmatrix}^{\top} \; .  \\
\intertext{Supposing that $x_{1}$ is the only state we can observe over time, our output is}
y(t) = x_{1}(t) \; . \label{eq:example_output}
\end{gather}

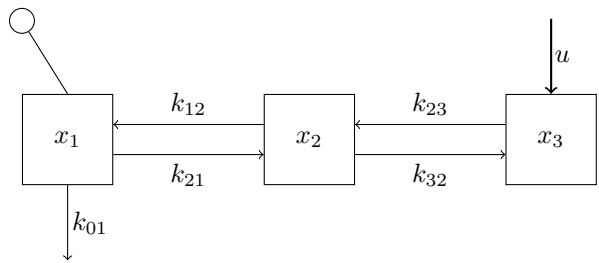
\begin{figure}[ht]
\centering
\begin{tikzpicture}
\node[draw,rectangle,minimum width=12mm, minimum height=12mm,black] (x2) at (0,0){$x_{2}$};
\node[draw,rectangle,minimum width=12mm, minimum height=12mm,black, left=2cm of x2] (x1) {$x_{1}$};
\node[draw,rectangle,minimum width=12mm, minimum height=12mm,black, right=2cm of x2] (x3) {$x_{3}$};
\draw[->] ([yshift=-2mm] x1.east) -- ([yshift=-2mm]  x2.west) node[pos=0.5, below]{$k_{21}$};
\draw[<-] ([yshift=2mm] x1.east) -- ([yshift=2mm] x2.west) node[pos=0.5, above]{$k_{12}$};
\draw[->] ([yshift=-2mm] x2.east) -- ([yshift=-2mm]  x3.west) node[pos=0.5, below]{$k_{32}$};
\draw[<-] ([yshift=2mm] x2.east) -- ([yshift=2mm] x3.west) node[pos=0.5, above]{$k_{23}$};
\draw[->] (x1.south) -- +(0,-1) node[pos=0.5, xshift=3mm]{$k_{01}$};
\draw[<-, thick] (x3.north) -- +(0,1) node[pos=0.5, xshift=1.5mm]{$u$};
\node[draw,radius=3mm, circle, above = 8mm of x1.north west] (x1obs) {};
\draw[] (x1.north) -- (x1obs){};
\end{tikzpicture}
\caption{A compartmental diagram of the chemical system as modelled by the representative model shown in  \eqref{eq:s1_example}--\eqref{eq:example_output}. Matter in the compartments representing the chemical species $x_{1}$, $x_{2}$, and $x_{3}$, is 
transferred between compartments. Matter is lost from the $x_{1}$ compartment to the environment and this
compartment is observed. Input $u$ delivers mass to the $x_{3}$ compartment.}  
\label{fig:comp_diagram} 
\end{figure}

We may represent this ``single-input single-output'' (SISO) structure by a compartmental diagram, as in Figure~\ref{fig:comp_diagram}.
Squares represent distinct chemical species, thin arrows show the conversion of mass to other forms, or excretion from the system.
The rates of conversion or excretion are determined by the product of the associated parameter and the state variable at the source of the arrow. The thick arrow shows an input, and the circle linked to $x_{1}$ indicates that this compartment is observed.
More specifically,  Figure~\ref{fig:comp_diagram}, \eqref{eq:s1_example}, and \eqref{eq:example_output} illustrate the representative model of a 
controlled (due to the input $u$) compartmental (mass is conserved) linear (describing the manner in which the states and input appear) time-invariant (coefficients of the input and states are constants) state-space structure.\footnote{We will treat classes of structures more formally in Sect.~\ref{s:prelims}.}

At this juncture, establishing some conventions will aid our further discussion of structures.

\begin{convention} \label{conv:gen_structure}
When discussing features of a structure $M$, we represent its associated parameter
space with $\Theta$, which we may specify more particularly as necessary. Given arbitrary parameter vector $\bth \in \Theta$, we shall always use $M(\bth)$ to 
represent $M$'s representative system. When considering some specific parameter vector, say $\boldsymbol{\alpha}$, we shall 
represent the associated model by $M(\boldsymbol{\alpha})$, which we will understand to be completely specified.
\end{convention}

\begin{convention} \label{conv:description}
When we apply some descriptors (e.g. controlled compartmental linear time-invariant state-space) to either a structure's representative system (as in the example above) or to a structure, these descriptors transfer to the other. The descriptors also apply to all systems in the structure, except for possibly degenerate systems associated with a subset of parameter space of measure zero.
\end{convention}

Convention~\ref{conv:description} foreshadows a case where some small number of models in a structure may have different properties to others. We may account for this complication in a manner that assists our intended analysis of structures.

\begin{property}
Given structure $M$ with parameter set $\Theta$, a property of $M$ is generic if it 
holds ``almost everywhere'' in $\Theta$. That is, we allow that the property may not hold on some subset(s) of $\Theta$
of measure zero.
\end{property}

Having specified a structure for a physical system, we may expect it to contain some model which will 
encapsulate the system's features of interest, and provide insights into aspects of the system's behaviour.
For example, we may hope to achieve objectives, such as to accurately:
\begin{description}
\item[O1] predict system outputs at unobserved times within the time range for which we have data,
\item[O2] estimate the time course of states,
\item[O3] anticipate system behaviour in situations for which we do not have data, such as under a proposed change in experimental or environmental conditions,
\item[O4] compare the effects of a range of proposed actions on the system, allowing us to discern which actions have the potential to produce beneficial results.
\end{description}

We can only hope to consistently gain such insights if our modelling effort provides reliable predictions. Yet, features of an assumed structure may make this challenging, or impossible. As such, we can benefit from interrogating structures in advance of their use to ascertain their suitability. 

To explain further, we may expect to arrive at a particular model in $M$ that we can use for prediction
after using data to estimate our parameter vector in a process of ``parameter identification'' (PI). 
In essence, PI uses some objective function to quantify the goodness-of-fit of predictions 
made for some $\boldsymbol{\alpha} \in \Theta$ to data, and  an algorithm that searches through $\Theta$ to improve
upon this as much as possible. The goal is to determine those parameter vectors which optimise the objective function. Suppose that there is a ``true'' (unknown) parameter vector $\bth^* \in \Theta$ such that $M(\bth^*)$ reproduces the actual dynamics of our physical system, including that relating to any unobservable states. As data is typically sparse and subject to noise, 
whilst we expect that we cannot exactly recover $\bth^*$, we intend that PI can obtain a good approximation to it.

This ambition is frustrated when the value of the objective function is virtually constant over some region of parameter space. Upon encountering such a region, a search algorithm is unable to find a search direction that will improve the objective function's value. This may lead to an unsatisfactory result. For example, the PI process may terminate without returning any parameter estimate.

 Alternatively, PI's results may defy interpretation. Suppose PI returns multiple feasible, equally valid estimates of $\bth^*$. If we lack further constraints on the elements of $\bth^*$ 
(e.g. relative sizes), we cannot discern which of the alternative estimates to use as our approximation.  

This state of affairs may not matter if our only concern is O1, or we do not need to specifically know 
$\bth^*$.
 However, suppose that using $M$ with alternative parameter estimates yields substantially different results for
 outcomes O2--O4. Then, we cannot confidently use $M$ for prediction.

Cox and Huber \cite{Cox_Symmetry_2007} provided one example of such an unsatisfactory outcome. The authors
showed that two parameter vectors returned by PI lead to equally good predictions of the observed time series of counts of malignant cancer cells in a patient, yet produce substantially different counts for 
the time after an ``intervention''--- a reduction in the carcinogenic components to which the patient is exposed.

PI may fail to uniquely estimate a parameter vector due an inherent property of $M$.  As such, our non-uniqueness problem 
is independent of the amount and quality of data we have. That is, improvements in the volume of data or accuracy of
its measurement cannot resolve the problem.

We expect to anticipate the non-uniqueness of parameter estimates when scrutiny of our structure
shows that it is not structurally globally identifiable (SGI).\footnote{The literature has various alternative terms for SGI, some of which
may be equivalent only under particular conditions.  
For two examples, Audoly et al. \cite{Nonlin_ID_Audoly}, used ``structurally a priori identifiable'', where ``a priori'' emphasises that one 
can test a structure in advance of data collection. Godfrey \cite{Godfrey_book} favoured ``deterministic identifiability'' in discussing
 compartmental models, for reasons relating to the degree of a priori knowledge of a system and the dependence of the result of 
 testing on the combination of inputs. We will consider this second matter in Section~\ref{s:controlled_LTI_ID}.} The concept was first 
formalised for state-space structures in Bellman and {\AA}ström \cite{Bellman_id} with reference to compartmental 
 structures similar to that shown in Figure~\ref{fig:comp_diagram}.

One tests a structure to determine whether or not it is SGI in an idealised framework.

\begin{convention} \label{conv:ID_framework}
The framework employed in testing a structure $M$ for SGI is defined by assumptions including:
\begin{itemize}
\item the structure correctly represents our physical system,
\item a record of error-free data that is infinite in extent is available,
\item and others that may be particular to the assumed class of structure, or testing method. 
\end{itemize}
Some methods, e.g. those employing similarity transforms \cite{Walter_Unidentifiable_1981} or Markov and initial parameters, \cite{van_den_Hof_95}, are only applicable when $M$ is ``generically minimal''.
 That is, for almost all $\bth \in \Theta$ we cannot reduce $M(\bth)$ to a system of fewer states that produces
  an identical output function.
\end{convention}

The test aims to discern whether or not it is possible for PI applied to idealised data to only return the true vector $\bth^*$, 
for almost all $\bth^* \in \Theta$. The test result is definitive in this case.

Suppose that structure $M$ is classified as SGI. Then, it may be possible for PI applied to actual (limited in extent, noisy) data to return a unique estimate for $\bth^*$, but this is not guaranteed. As such, we can only consider an SGI model as possibly useful for prediction. 
Still, the value of knowing that $M$ is SGI is the assurance that we are not almost certain to fail in our 
objective before we commence our study.
Alternatively, it is extremely unlikely that PI applied to a non-SGI model and actual data will return a unique estimate of 
$\bth^*$. In this case, we should not immediately proceed to make predictions following PI. Instead, we may seek to propagate parameter uncertainty through our structure so as to produce a range of predictions, allowing us to quantify prediction uncertainty. From this we may judge whether or not we can obtain sufficiently useful predictions for our purposes. 

Aside from merely encouraging caution, the result of testing structure $M$ for structural global identifiability\footnote{In the interests of brevity, henceforth we use SGI as a shorthand for this noun, in addition to the adjective used earlier, expecting that the reader can infer the meaning from context.} can deliver useful insights. 
The test result may allow us to distinguish between individual parameters we may estimate uniquely, and those we cannot. 

Further, awareness  that a structure is not SGI can assist in correcting the problem. The test may allow us to recognise those parameter combinations which PI may return uniquely. This knowledge may guide reparameterisation of $M(\bth)$ so as
to produce the representative system of a new structure that is SGI. Additionally, having learned that $M$ is not SGI, one can examine whether it is possible that modifying $M$ (e.g. holding some parameters constant), or the combination of
$M$ and planned data collection (e.g. supposing that an additional variable is measured, and rewriting $M$ to include this as
another output), will remedy this.
Thus, we can treat the process of testing a structure for SGI as an iterative process. We can detect a structure's undesirable features ahead of data collection, address them, test the revised structure, and continue this process until the structure is satisfactory.

Analytical inspection of (in particular, more complex) structures to anticipate the uniqueness or otherwise of parameter estimates is often not straightforward. The difficulties of testing a structure for SGI, as well as how the results of PI applied to real data can be worse than that predicted by theory, have encouraged numerical approaches to the task. (See \cite[Chapter~8]{Godfrey_book} for an introduction.) Broadly, approaches seeking to demonstrate ``numerical'' (or ``practical'') identifiability are based on assuming some number of parameter vectors; using each of these with the structure to simulate data at a limited number of observation times, or under a limited number of conditions (e.g. applied inputs or values of experimental variables), or subject to noise, or some combination of these; conducting PI; and investigating the features of parameter estimates to determine if these adequately approximate assumed values. 

Testing a structure for numerical identifiability may determine when PI is unlikely to yield accurate results.  However, unlike analytical scrutiny, these investigations may not provide clear guidance on how to remedy the problem.

In this paper we will provide an introduction to the testing of (state-space) structures for SGI.  
There are a variety of testing methods available (see, for example, \cite{DiStefano_Dynamic_2015}) although many are 
not an ideal means of introducing the field of identifiability analysis.
As such, we intend that our choices of testing method and examples will allow us to illustrate some important issues without having to encounter unnecessary algebraic and conceptual complexity.

In choosing example structures, we have limited ourselves to a class which are linear in the state variables,
as demonstrated in the representative model given in \eqref{eq:s1_example}--\eqref{eq:example_output} . We further restrict these to compartmental structures. Given these choices, the ``Transfer Function Approach'' (TFA, see for example \cite{Ciobelli_Parameter_1980}), which makes use of features of the Laplace transform of a structure's output function,\footnote{For this reason, the approach is also known as the ``Laplace transform method'', as seen in \cite[Chapter~6]{Godfrey_book}.}  is appropriate for our purposes.
Although one of the older testing methods, it is still included in relatively recent texts presenting a range of methods (e.g. \cite{DiStefano_Dynamic_2015}), and:
\begin{enumerate}
\item is conceptually rather more straightforward than other methods, 
\item has the unusual distinction of being applicable to a structure that is not generically minimal, and
\item is unambiguously appropriate for compartmental structures.
\end{enumerate}

To explain the significance of Points 2 and 3, we note that a general linear state-space structure
may be judged as generically minimal as a consequence of having the generic properties of controllability 
and observability. The conditions used in deciding this are appropriate for linear systems--- these have
a state space which is a vector space. However, the state space of a positive linear system is a polyhedral cone,
and so it does not seem appropriate to treat these as we would a general linear system. 

Certain authors have sought to highlight differences between features of linear and linear positive systems. In 
the context of discrete-time systems, Benvenuti and Farina sought to show 
\begin{quote}
\ldots that the minimality problem for positive linear systems is inherently different from that of ordinary linear systems \ldots'' (\cite[Page~219]{Benvenuti_Minimal_2003}).
\end{quote}
Whyte \cite[Chapter 3, Section~5.2]{Whyte_PhD_2016} considered some of the literature's perspectives on controllability of linear state-space systems. Briefly, the origins of the area related to linear ``structured'' systems (see Poljak \cite{Poljak_TVstruc}) which are generally distinct from linear compartmental systems (a type of ``descriptor'' system;
 see Yamada and Luenberger \cite{Yamada_aut_con85}). This lead to suspicions that it may not always be inappropriate to test a linear compartmental structure for generic minimality using the machinery designed for general linear structures.
By choosing to use the TFA in analysing a structure, Point 2 allows us to avoid this potential issue.

Further, the TFA has shown promise in the analysis of structures of linear switching systems (LSSs) (Whyte~\cite{Whyte_PhD_2016, Whyte_Inferring_2013}). Structures of switching systems (especially those which evolve in continuous time) are largely neglected in the literature. Yet methods under development may assist in the scrutiny of structures used to model epidemics, such as where an intervention causes an abrupt change in some parameter values.

Discussions at a recent workshop ``Identifiability problems in systems biology'' held at the American Institute of Mathematics (\cite{AIM_Sys_Bio2019}) highlighted a degree of inconsistency in certain key definitions used in the field of identifiability analysis. As such, here we will draw on efforts to propose transparent and coherent
definitions in the analysis of uncontrolled structures (Whyte~\cite{Whyte_PhD_2016, Whyte_Inferring_2013})
in suggesting equivalent definitions for controlled structures.

The remainder of this paper is organised as follows. In Section~\ref{s:prelims} we present some preliminary material and 
introduce certain classes of structures that aid us in presenting the TFA. In Section~\ref{s:ID_outline} we outline the general 
theory of testing an uncontrolled structure for SGI, particularise this to uncontrolled linear time-invariant (LTI) state-space structures, and consider an
example. Section~\ref{s:controlled_LTI_ID} proceeds similarly for controlled LTI state-space structures, where we draw an important distinction between testing approaches based on how much information we are able to elicit from our structure. Finally, in Sect.~\ref{s:conclusions} we summarise some concepts in the testing of structures and offer some concluding remarks. 

We conclude this section by establishing notation.

\subsection{Notation}
The field of real numbers is denoted by $\mathbb{R}$. The subset of $\mathbb{R}$ containing only positive (non-negative) values is denoted by $\mathbb{R}_{+}$ ($\bar{\mathbb{R}}_{+} $). The natural numbers $\{1,2,3,\ldots \}$ are denoted by $\mathbb{N}$, and we define $\mathbb{N}_{0} \triangleq \mathbb{N} \cup \{ 0 \}$. 

The field of complex numbers is denoted by $\mathbb{C}$. The real part of $z \in \mathbb{C}$ is denoted by $\mbox{Re}(z)$.
Given some $a \in \mathbb{R}$, a useful set for the following discussion is 
\begin{align}
{\rm H}_a \triangleq \{  s \in \mathbb{C} \big| {\rm Re}(s) > a \} \, . \label{eq:H_set}
\end{align}
 
We use a bold lower-case (upper-case) symbol such as  $\vec{a}$ ($\vec{A}$) to denote a vector (matrix), and a superscript ${\top}$ associated with any such object indicates its transpose.  Given vector $\vec{a}$,  $\dot{\bf a}$ denotes its derivative with respect to time.
To specify the $(i,j)$-th element of $\vec{A}$ we may use $a_{i,j}$, or prefer the simplicity of $(\vec{A})_{i,j}$ when $\vec{A}$ is a product of terms.
For $n \in \mathbb{N}$, we use ${\rm diag}(a_{1}, a_{2}, \ldots, a_{n})$ to denote the square diagonal matrix 
having $a_{1},\ldots,a_{n}$ on the main diagonal and zeros elsewhere. 
A special diagonal matrix is the ($n \times n$) identity matrix $\vec{I}_{n} \in \mathbb{R}^{n \times n}$, having a main diagonal of $n$ 1s.

 Given field $\mathbb{F}$ and some indeterminate $w$, $\mathbb{F}(w)$ denotes the field of
rational functions in $w$ over $\mathbb{F}$.
 Given $a,b\in \mathbb{N}_{0}$ and $\mathbb{F}$, we use $\mathbb{F}^{a \times b}$ to denote the
set of matrices of $a$ rows and $b$ columns having elements in $\mathbb{F}$. 
When at least one of $a$ or $b$ is zero, it is convenient to have $\mathbb{F}^{a \times b}$  represent a set of ``empty matrices'', and we can disregard any matrix in this set as it arises.

\section{Preliminaries}   \label{s:prelims}

In this section we will define certain classes of structures, and present an overview of some useful properties, in preparation for a discussion of how we may test these structures for SGI.

We will aim to illustrate the features of systems by introducing sufficient systems theory, beginning with some conventions. 
Suppose we have a set of input values $U$, a set of output values $Y$, and a time set $T \subseteq \bar{\mathbb{R}}_{+}$. 
Let ${\cal U}$ denote a set of input functions such that for 
$u \in {\cal U}, u:T \rightarrow U^{T}:t \mapsto u(t) \in U$. That is,  ${\cal U}$ is a set of input functions taking values in the set $U$. 
Similarly, let ${\cal Y}$ denote a set of functions such that for $y \in {\cal Y}, y:T \rightarrow Y^{T}:t \mapsto y(t) \in Y$. That is,  ${\cal Y}$ is a set of output functions taking values in a set~$Y$. Finally, let $\zeta$ denote an ``input-output'' map from ${\cal U}$ to ${\cal Y}$. We use these definitions in presenting a general type of system in Definition~\ref{defn:input_output}. From this 
we may obtain other system types by imposing suitable conditions.

\begin{definition} \label{defn:input_output}
An {\bf input-output system} on time set $T$ is a triple $({\cal U, Y},\zeta)$.
\end{definition}

Contained within the input-output systems are the state-space systems, which are of particular interest to us here.
To aid our discussion of these, given some time set $T$ we define the set
\begin{align}
T^{2}_{+} \triangleq \big\{ ( t_{2},t_{1}); \ t_{2} \ge t_{1}, \ t_{1},t_{2} \in T \big\} \; .  
\label{eq:Tplus2}
\end{align}

\subsection{State-space structures}

 In the following definitions and discussion we draw on Whyte~\cite[Section~3.4]{Whyte_PhD_2016}, which was informed by Caines \cite[Appendix 2]{Caines_systems}).

\begin{definition}[Adapted from Whyte~{\cite[Definition 3.8]{Whyte_PhD_2016}}] \label{defn:state_space}
A {\bf state-space system} $\Sigma$ is a quintuple 
$({\cal U}, X, {\cal Y}, \Phi, \eta)$ where
\begin{itemize}
\item ${\cal U}$ is a set of input functions.
\item $X$ is a set, called the state-space of $\Sigma$, with elements called states.
\item ${\cal Y}$ is a set of output functions.
\item $\Phi(\cdotp, \cdotp,\cdotp,\cdotp)$ is the state transition function, which maps 
$ T^{2}_{+} \times X \times {\cal U}$ into $X$. \\
 To illustrate this, consider time interval
$T \subseteq \bar{\mathbb{R}}_+$ with $t_0 \triangleq \inf T $. Suppose $\Sigma$ is subject to input function $u \in \mathcal{U}$. Further, suppose that at $t=t_0$ we have that $x_0 \in X$ is the initial state of $\Sigma$.
Then, for $(t,t_0) \in T_{+}^2$, $\Phi(t,t_0,x_0,u)$ determines the state of $\Sigma$ as a consequence of time $t$, $x_0$, and $u$.
Under these conditions, we may concisely refer to $\Phi(t,t_0,x_0,u)$ as the state of $\Sigma$ at time $t$. 

\item $\eta(\cdotp,\cdotp,\cdotp)$ is the output map,  which maps $T \times X \times {\cal U}$ into $Y$. \\
That is, at some time $t \in T$, $\eta$ determines the output vector that results from three inputs: $t$, the state of $\Sigma$ at that time, and the input $u$.
\end{itemize}
Further, the following four properties hold:
\renewcommand{\labelenumi}{SS\arabic{enumi}:}
\begin{enumerate}
\item The Identity Property of $\Phi$ 
\begin{align*}
\Phi(t,t,x,u)=x, \mbox{ for all } t\in T, \ x\in X  \mbox{ and } u \in {\cal U} \; .
\end{align*}
That is, suppose the state of $\Sigma$ at time $t$ is $x$. Then, if no time has elapsed from $t$, $\Phi$ does not move the state away from $x$.
\item The Nonanticipative Property of $\Phi$ \\
Suppose we have any $u_{1}, u_{2} \in {\cal U}$ such that these functions are identical on time interval $[t_{0},t_{1}]$,
where $(t_{1},t_{0}) \in T^{2}_{+} \subset \mathbb{R}^{2}_{+}$.
Then, for all $x \in X$ we have 
\begin{align*}
\Phi(t_{1},t_{0},x,u_{1}) = \Phi(t_{1},t_{0},x,u_{2}) \; .
\end{align*}
To explain this, suppose the state of $\Sigma$ at time $t_0$ is some $x \in X$. The Nonanticipative Property of $\Phi$ means that $\Sigma$ reaches the same state at time $t_1$ for $\Phi$ subject to either $u_1$ or $u_2$. Equivalently, 
differences between $u_1$ and $u_2$ for any time greater than $t_1$ do not influence the evolution of the state
of $\Sigma$ on $[t_0,t_1]$ under $\Phi$.
\item The Semigroup Property of $\Phi$  \\
For all $(t_{1},t_{0}), (t_{2},t_{1}) \in T^{2}_{+}$, $x \in X$,
and $u \in {\cal U}$,
\begin{align*}
\Phi(t_{2},t_{0},x,u) =
\Phi \big(t_{2},t_{1},\Phi(t_{1},t_{0},x,u),u \big) \; . 
\end{align*}
To explain, suppose we have system $\Sigma$ with initial state $x$ at time $t_0$ and input $u$. 
Suppose $\Phi$ acts on time interval $[t_0,t_1]$ resulting in some particular state (say $x_1 \triangleq \Phi(t_{1},t_{0},x,u)$) at $t_1$. Suppose then $\Phi$ uses $x_1$ as an initial state for evolving the state of $\Sigma$ on $[t_1,t_2]$, 
resulting in a particular state (say $x_2 \triangleq \Phi \big(t_{2},t_{1},\Phi(t_{1},t_{0},x,u),u \big)$) at $t_2$. 
Due to the Semigroup Property of $\Phi$, system $\Sigma$ also reaches state $x_2$ at $t_2$ if $\Phi$ is used to evolve the state on $[t_0,t_2]$.
\item The Instantaneous Output Map $\eta$ \\
For all $ x \in X$, $u \in {\cal U}$, $ (t,t_{0}) \in
T^{2}_{+}$, the function $y:T \rightarrow Y$ defined via 
\begin{align*}
y(t) = \eta \big(t,\Phi(t,t_{0},x,u),u(t) \big)
\end{align*}
is a segment of a function in ${\cal Y}$. \\
That is, we can use $\eta$ to define the instantaneous output of $\Sigma$ at current time $t$ through $t$, the 
state of $\Sigma$ at time $t$ ($\Phi(t,t_{0},x,u)$) and the value of the input at time $t$ ($u(t)$).
This property is useful as $y$ provides a simpler means of illustrating the output of $\Sigma$ than does $\eta$ when we wish to introduce particular system types.
\end{enumerate}
\end{definition}

We will now illustrate some useful classes of continuous-time state-space structures, beginning with a general type.
Henceforth we consider spaces for states, inputs, and outputs of $X \subseteq \mathbb{R}^{n}$, $U \subseteq \mathbb{R}^{m}$, and  $Y\subseteq \mathbb{R}^{k}$, respectively, where accordingly indices $n, m,k \in \mathbb{N}$ 
determine the dimensions of our state, input, and output vectors. For arbitrary parameter vector 
$\bth \in \Theta$, and input ${\bf u} \in \mathcal{U}$,  at time $t \in T$ a controlled state-space structure $M$ has representative system $M(\bth)$ of the general form:
\begin{align}
\begin{gathered}
\dot{ {\bf x} } (t;\bth)  = {\bf f}({\bf x },{\bf u }, t;\bth), \quad {\bf x}(0; \bth ) = {\bf x}_{0}(\bth) \; ,  \\
 {\bf y} (t; \bth)  = {\bf g}({\bf x },{\bf u }, t;\bth) \; ,
\end{gathered} \label{eq:general_controlled} 
\end{align}
where ${\bf f}$ and ${\bf g}$ satisfy the relevant properties SS1--SS4 of Definition~\ref{defn:state_space}.

A subtype of the controlled state-space structures are an uncontrolled class, lacking inputs.
If an uncontrolled state-space structure has indices for the state and output spaces of $n$ and $k$ respectively, then a representative model is similar to \eqref{eq:general_controlled}:
\begin{gather}
\begin{gathered}
\dot{ {\bf x} } (t ;\bth)  = {\bf f}({\bf x }, t;\bth), \quad {\bf x}(0; \bth ) = {\bf x}_{0}(\bth) \; ,  \\
\dot{ {\bf y} } (t; \bth)  = {\bf g}({\bf x }, t; \bth) \; .
\end{gathered} \label{eq:general_uncontrolled} 
\end{gather}

We will now introduce a particular class of the general state-space structures described above--- that of linear time-invariant (LTI) structures. 
An LTI structure has a representative system that is particular form of \eqref{eq:general_controlled}. 
We will use specific examples of LTI structures to illustrate the testing of a structure for SGI in Sections~\ref{s:ID_outline} and~\ref{s:controlled_LTI_ID}.

\subsection{Continuous-time linear, time-invariant structures}

The following definitions are adapted from Whyte~\cite[Definition~3.21]{Whyte_PhD_2016}, which drew on concepts 
from van~den~Hof~\cite{van_den_Hof_95}. 

\begin{definition} \label{defn:LTI_structure}
Given indices $n,m,k \in \mathbb{N}$, a {\bf controlled continuous-time linear time-invariant state-space structure} (or, more briefly, {\bf an LTI structure}) $M$ has state, input, and output spaces $X = \mathbb{R}^{n} $, $U = \mathbb{R}^{m}$, and $Y = \mathbb{R}^{k}$, respectively. 
For parameter set $\Theta \subseteq \mathbb{R}^{p}$ ($p \in \mathbb{N}$), $M$ has mappings 
\begin{equation}
{\bf A}: \Theta \rightarrow \mathbb{R}^{n\times n} \;, \quad
{\bf B}: \Theta \rightarrow \mathbb{R}^{n\times m} \; , \quad
{\bf C}: \Theta \rightarrow \mathbb{R}^{k\times n} \; , \quad
{\bf x_{0} }: \Theta \rightarrow \mathbb{R}^{n} \; ,  \label{eq:ss_matrices}
\end{equation}
where the particular pattern of non-zero elements in the ``system matrices'' shown in \eqref{eq:ss_matrices} defines $M$.
More specifically, mappings in \eqref{eq:ss_matrices} dictate the relationships between state variables ${\bf x}$, 
inputs ${\bf u}$, and outputs ${\bf y}$ for all times $t \in T \subseteq \mathbb{R}_{+}$. Thus,
for arbitrary $\bth \in \Theta$, $M$'s representative system $M(\bth)$  has the form
\begin{gather}
\dot{ {\bf x} } (t, {\bf u} ; \bth ) = {\bf A}(\bth) \cdot {\bf x} (t, {\bf u} ; \bth) + {\bf B}(\bth) \cdot {\bf u} (t) \; , 
\quad {\bf x}(0; \bth) = {\bf x}_{0}(\bth) \; ,    \label{eq:controlled_LTI_state}   \\f
  y(t;\bth) = {\bf C }(\bth) \cdot {\bf x} (t;\bth) \; .  \label{eq:controlled_LTI}
\end{gather}

Defining
\begin{gather}
 L\Sigma P(n,m,k) \triangleq \mathbb{R}^{n\times n} \times
\mathbb{R}^{n\times m} \times \mathbb{R}^{k \times n} \times \mathbb{R}^{n} \; , \label{eq:system_matrices_set}  \\
\intertext{then}
SL \Sigma P(n,m,k) \triangleq \left\{ \left. \Big( {\bf A}( \boldsymbol{\theta}),
{\bf B}( \boldsymbol{\theta}),{\bf C}( \boldsymbol{\theta}), {\bf x_{0}}( \boldsymbol{\theta}) \Big)  \in L \Sigma P(n,m,k)  \right| \boldsymbol{\theta} \in \Theta  \right\}  \label{eq:structure_matrices_set}
\end{gather}
is the set of system matrices associated with systems in $M$. Thus, we may consider the matrices of a particular system in $M$ 
as obtained by the parameterisation map $ f: \Theta \rightarrow SL \Sigma P(n,m,k)$ such that
\begin{align*}
f({\bf \boldsymbol{\theta} }) =  \Big( {\bf A}( \boldsymbol{\theta}),
{\bf B}( \boldsymbol{\theta}),{\bf C}( \boldsymbol{\theta}),
{\bf x_{0}}( \boldsymbol{\theta}) \Big).
\end{align*}

Together, the matrices and vector defined by \eqref{eq:ss_matrices} and the indices $n$, $m$, and $k$, are the {\bf system parameters} of $M(\bth)$.  

We may consider an {\bfseries uncontrolled LTI structure} having indices $n,k \in \mathbb{N}$ 
as a form of controlled LTI structure having $n,m,k \in \mathbb{N}_{0}$ by setting $m =0$. As such, 
systems in the uncontrolled structure have $X = \mathbb{R}^{n}$ and $Y = \mathbb{R}^{k}$. 
By omitting the empty matrix ${\bf B}$ from \eqref{eq:controlled_LTI_state} we obtain the form of the uncontrolled structure's representative system:
\begin{gather}
\dot{{\bf x} } (t;\bth)  = {\bf A}(\bth)\cdot {\bf x} (t;\bth), \quad {\bf x}(0;\bth) = {\bf x}_{0}(\bth) \; , 
\label{eq:LTI_ode_state} \\
  y(t;\bth) = {\bf C }(\bth) \cdot {\bf x} (t;\bth) \; ,  \label{eq:LTI_y}
\end{gather}
where the system matrices are ${\bf A} \in \mathbb{R}^{n \times n}$, ${\bf C} \in \mathbb{R}^{k \times n}$,
and $\vec{x_{0}} \in \mathbb{R}^{n}$.

As a notational convenience, we allow sets defined in \eqref{eq:system_matrices_set} and \eqref{eq:structure_matrices_set} to apply to this context, where $L\Sigma P(n,0,k)$ and $SL\Sigma P(n,0,k)$ 
are understood as neglecting the irrelevant $\vec{B}$.
\end{definition}

In modelling biological systems, we may employ a subclass of the LTI state-space structures in which systems have states, inputs, and outputs subject to constraints informed by physical considerations. This, in turn, imposes conditions on the structure's system matrices. Our summary of the conditions in the 
following definition is informed by the treatment of compartmental LTI systems given in van~den~Hof~\cite{van_den_Hof_95}.

\begin{definition}[Classes of LTI state-space structures] \label{def:LTIpos_structure}
A {\bf positive LTI state-space structure} with indices $n,m,k \in \mathbb{N}$ is an LTI state-space structure after Definition~\ref{defn:LTI_structure},
having representative
system of the form given in \eqref{eq:controlled_LTI_state} and \eqref{eq:controlled_LTI},   
where states, outputs, and inputs are restricted to non-negative values.  That is, the structure has $X = \bar{\mathbb{R}}^n_+$, $U = \bar{\mathbb{R}}^m_+$, and $Y = \bar{\mathbb{R}}^k_+$. 

A {\bf compartmental LTI structure} with indices $n,m,k \in \mathbb{N}$ is a positive LTI state-space structure for which systems in the structure have  system matrices subject to ``conservation of mass'' conditions:
\begin{itemize}
\item all elements of ${\bf B}$ and ${\bf C}$ are non-negative, and 
\item for  ${\bf A} =(a_{i,j})_{i,j=1,\ldots,n}$,
\begin{align}
\begin{alignedat}{2}
a_{ij} & \ge  0 \; , & \quad  &  i,j \in \left\{1,\ldots,n \right\}, \  i \ne j \; , \\
a_{ii} & \le  - \sum^{n}_{\substack{j = 1 \\ j \ne i}}  a_{ji} \; , & \quad & i  \in \left\{1,\ldots,n \right\} \; .
\end{alignedat} \label{eq:mass_cons_A}
\end{align}
\end{itemize}
An {\bf uncontrolled positive LTI structure} or an {\bf uncontrolled compartmental LTI structure} 
with indices $n,k$ belongs to a subclass of the corresponding class of controlled LTI structures
with indices $n,k,m$. The relationship between the controlled and uncontrolled forms is as for that between LTI structures and uncontrolled LTI structures presented in Definition~\ref{defn:LTI_structure}.
The representative system of any such uncontrolled structure has the form outlined in 
\eqref{eq:LTI_ode_state} and \eqref{eq:LTI_y}, subject to appropriate restrictions on state and output spaces
$X$ and $Y$.
\end{definition}

We shall now consider some properties of controlled LTI structures which will inform our testing of these structures for SGI
subsequently.

\subsection{Features of the states and outputs of a controlled LTI structure}

A consideration of some features of the states and outputs of  LTI structures here will allow us to appreciate the utility of the
TFA in testing such a structure for SGI in Section~\ref{s:ID_outline}.

\subsubsection{The time course of states and outputs} \label{sss:time_course}
In this discussion we adapt the treatment of uncontrolled LTI systems given in Whyte~\cite[Chapter~3]{Whyte_PhD_2016} and combine this with insights from Seber and Wild \cite[Chapter~8]{Seber_Wild_2003}. In this subsection, in the interests of brevity, we
we will neglect the dependence of systems on $\bth$.

Let us consider a structure defined by system matrices in $SL \Sigma P(n,m,k)$ (recall \eqref{eq:system_matrices_set}),
where we assume the structure is defined on time set $T = \bar{\mathbb{R}}_{+}$.
Recall that states evolve according to an ODE system as in \eqref{eq:controlled_LTI_state}. Given state space 
$X = \mathbb{R}^{n}$, the solution for state vector $\vec{x}(t)$ depends on the 
matrix exponential $e^{\vec{A}t} \in  \mathbb{R}^{n \times n}$ through
\begin{align}
\vec{x}(t) = \E^{\vec{A} t} \vec{x_{0}} + \int_{0}^{t} \E^{\vec{A}(t- t^{\prime})} \vec{B} \vec{u}(t^{\prime}) \D t^{\prime} \; ,
\label{eq:xt_soln}
\end{align}
provided that the integral exists. Assuming this existence, we may use \eqref{eq:LTI_y} and the convolution operator $\ast$
to express response as
\begin{align}
\vec{y}(t) = \vec{C} \E^{\vec{A}t} \vec{x_{0}} +  \vec{C} \E^{\vec{A}t} \vec{B} \ast \vec{u}(t) \; . \label{eq:alt_matrix_exp}
\end{align}

Let us presume a situation typical in the modelling of physical systems---that the elements of $\vec{A}$ are finite.
Let us suppose that the $n$ (finite and not necessarily distinct) eigenvalues of $\vec{A}$ are ordered from largest to smallest and labelled as $\lambda_{i}$, $i=1,\ldots,n$. In the interests of simplicity,  we also assume that $\vec{A}$ has $n$ linearly independent right eigenvectors $\vec{s}_{i}$, $i=1,\ldots,n$, where each is associated with the appropriate $\lambda_{i}$. 
We define ${\bf S} \in \mathbb{R}^{n \times n}$ as the matrix for which the $i$-th column is $\vec{s}_{i}$.
We may then employ a spectral decomposition $\vec{A} \equiv \vec{S \bm{\varLambda} S^{-1}}$,
where $\bm{\varLambda} = {\rm diag}(\lambda_{1}, \ldots , \lambda_{n})$. As a result, we may rewrite our matrix exponential:
 \begin{align}
 \E^{\vec{A}t} \equiv \vec{S} \E^{\bm{\varLambda} t } \vec{S}^{-1} \; , 
 \end{align}
noting that each element is a sum of (up to $n$) exponentials, with exponents drawn from $\lambda_{i}$ ($i=1,\ldots,n$).

With this in mind, let us turn our attention towards the terms 
$ \vec{C} \E^{\vec{A}t} \vec{x_{0}} \in  \mathbb{R}^{k \times 1}$ and $ \vec{C} \E^{\vec{A}t} \vec{B}  \in
\mathbb{R}^{k \times m} $ on the the right-hand side of \eqref{eq:alt_matrix_exp}.
 As $\vec{x_{0}}$ is a constant vector, and $\vec{B}$ and $\vec{C}$ are constant matrices, 
 then each element of $ \vec{C} \E^{\vec{A}t} \vec{x_{0}} $ and $ \vec{C} \E^{\vec{A}t} \vec{B}$ 
is also a sum of exponentials in $\lambda_{i}$ ($i=1,\ldots,n$). 

Suppose $\lambda_1$ has multiplicity $\mu \ge 1$. Hence, the largest possible dominant term in any of our sums of exponentials involves
$t^{\mu} e^{\lambda_1 t}$. 
Hence, there exist real constants $K > 0$ and $\lambda  > \lambda_1$ such that for all $t \in \bar{\mathbb{R}}_{+}$ we have 
\begin{align}
K \E^{\lambda t}  \ge \left\{ \begin{matrix}
\left| \left(   \vec{C} \E^{\vec{A}t} \vec{x_{0}}  \right)_{i,1}  \right| & i=1,\ldots, k \; , \\[+15pt]
\left| \left(   \vec{C} \E^{\vec{A}t} \vec{B}  \right)_{i,j} \right|  & 
\begin{aligned}[t]
i& =1,\ldots, k \; , \\ 
 j&=1,\ldots, m \; . 
 \end{aligned}
\end{matrix}   \right.
\label{eq:trans_matrix_elements_bound}
\end{align}

The existence of these bounds will prove important when we consider the application of the TFA to a LTI structure.
Towards this, we shall consider some features of the Laplace transform of the output of LTI structures.

\subsubsection{The Laplace transform of an LTI structure output function} \label{ss:LTI_LTy}

We recall the definition of the Laplace transform of a real-valued function.
\begin{definition}
Suppose some real-valued function $f$ is defined for all non-negative time. (That is,
$f: \bar{\mathbb{R}}_{+} \mapsto \mathbb{R} , \  t  \mapsto f(t)$.)
We represent the (unilateral) Laplace transform of $f$ with respect to the transform
variable $s \in \mathbb{C}$ by
\begin{equation*}
\mathcal{L} \{ f \} (s) \triangleq \int_{0}^{\infty} f(t)\cdot e^{-st} {\rm d}t \; ,
\end{equation*}
if this exists on some domain of convergence $\mathcal{D} \subset \mathbb{C}$.
\end{definition}

Let us consider a controlled LTI structure $S$ with parameter set $\Theta$, with a representative system 
$S(\bth)$, having the form shown in \eqref{eq:controlled_LTI_state} and \eqref{eq:controlled_LTI}.  
We assume system matrices belong to $SL \Sigma P(n,m,k)$ (recall \eqref{eq:structure_matrices_set}).
Suppose that given input $\vec{u}$,
$\mathcal{L}\{\vec{u}\}(s)$ exists.
In this case the Laplace transform of output ${\bf y}$ given ${\bf u}$ is\footnote{We note that others, such as Walter and Pronzato \cite[Chapter 2, Page 22]{Walter_book}, have considered such expressions. However, the notation employed
may make the description of transfer functions in testing a structure for SGI unnecessarily complicated. As such, 
we employ a simpler notation here.  We also include ${\bf x_{0}} $ in ${\bf V}$ (unlike say in the equivalent matrix ${\bf H_{2} }$ in \cite{Walter_book}), as otherwise the initial conditions do not feature in the test equations.}
\begin{align}
\mathcal{L} \{ { \bf y} (\cdot, {\bf u}; \bth ) \}(s;\bth) & = { \bf V }(s; \bth) +
 { \bf W }(s; \bth) \mathcal{L} \{ { \bf u } \}(s) \in \mathbb{R}(s)^{k \times 1} \; , \label{eq:LT_controlled_y} \\
\intertext{where}
 { \bf V } (s; \bth) & \triangleq {\bf C}(\bth) \big( s {\bf I}_{n} - {\bf A}(\bth) \big)^{-1} { \bf x_{0}} (\bth)
 \in \mathbb{R}(s)^{k \times 1} \;  , \label{eq:V} \\
{ \bf W}(s; \bth) & \triangleq {\bf C}(\bth) \big( s{\bf I}_{n} - {\bf A}(\bth) \big)^{-1} {\bf B}(\bth) 
 \in \mathbb{R}(s)^{k \times m} \; ,  \label{eq:W} 
\end{align}
and, owing to \eqref{eq:trans_matrix_elements_bound}, each element of $\vec{V}$ and $\vec{W}$  is defined for all  $ s \in H_{\lambda}$.

\begin{definition} \label{defn:unproc_TF}
We refer to $\vec{V}$ and $\vec{W}$ as ``transfer matrices'', and each element of these is a transfer 
function--- specifically, a rational function in $s$. We term any such element an {\it unprocessed transfer function}.
\end{definition}

\begin{property}
The degree of the denominator of any unprocessed transfer function 
in $\vec{V}$ or $\vec{W}$ is at most $n$. 
Similarly, if $S$ is a compartmental structure, the degree of the numerator of any transfer function is at most $n-1$. 
If we can cancel any factors in $s$ between the numerator and denominator of the transfer function (pole-zero cancellation), then we will obtain a degree for each of the numerator and denominator which is lower than previously.

Suppose that pole-zero cancellation occurs in each unprocessed transfer function in $\vec{V}$ and $\vec{W}$. Then,
$S$ is not generically minimal (recall Convention~\ref{conv:ID_framework}). 
\end{property}

When we have an uncontrolled LTI structure, \eqref{eq:LT_controlled_y} reduces to 
\begin{align}
\mathcal{L} \{ { \bf y} ( \cdot ; \bth ) \}(s) & = { \bf V }(s; \bth) \in  \mathbb{R}(s)^{k \times 1} \; , \label{eq:LT_uncontrolled_y} 
\end{align}
with ${\bf V}$ as in \eqref{eq:V}, and the discussion of matrix elements given above also applies.

We may now proceed to consider definitions and processes relating to structures and structural global
identifiability, informed by Convention~\ref{conv:ID_framework}. By way of introduction, we begin with the rather more straightforward matter of the testing of uncontrolled structures.
 
\section{Testing an uncontrolled structure for structural global identifiability} \label{s:ID_outline}

We will consider the testing of an uncontrolled structure for SGI following what we may call the ``classical'' approach 
originally outlined by Bellman and {\AA}ström \cite{Bellman_id}. We follow the treatment of \cite{Whyte_PhD_2016} 
which drew on aspects of Denis-Vidal and Joly-Blanchard  \cite{Denis_auto_equiv04}. In essence, we judge a structure as SGI (or otherwise) with  reference to the solution set of test equations. 

\begin{definition} \label{def:ID}
Suppose we have a structure of uncontrolled state-space systems $M$, having parameter set $\Theta$ (an open
subset of $\mathbb{R}^{p}$, $p \in \mathbb{N}$), and time set $T \subseteq [0, \infty)$.
For some unspecified $\bth \in \Theta$, $M$ has representative model $M(\bth)$, which has 
state function $\vec{x}(\cdot; \bth) \in \mathbb{R}^{n}$ and output ${\bf y}(\cdot; \bth) \in \mathbb{R}^{k} $  (recall \eqref{eq:general_uncontrolled}). 
Suppose that systems in $M$ satisfy conditions:
\begin{enumerate}
\item The functions $\vec{f}(\vec{x}, \cdot; \bth)$ and $\vec{g}(\vec{x}, \cdot; \bth)$  are real and analytic for every $\bth \in \Theta$
on $\mathcal{S}$ (a connected open subset of $\mathbb{R}^{n}$  such that $\vec{x}(t; \bth) \in \mathcal{S}$ for every
$t \in [0, \tau]$, $\tau >0$).
\item $\vec{f}(\vec{x_{0}}(\bth); \bth) \neq \vec{0}$ for almost all $\bth \in \Theta$. 
\end{enumerate}
Then, for some finite time $\tau >0$, we consider the set
\begin{align}
\displaystyle {\mathcal I}(M) \triangleq \left\{ \boldsymbol{\theta^{'}} \in \Theta: 
{\bf y}(t; \boldsymbol{\theta^{'}}) = {\bf y}(t; \bth)  \quad \forall t \in [0,\tau]  \right\} \; . \label{eq:ID_def}
\end{align}
If, for almost all $\bth \in \Theta$:
\begin{description}
\item  $   {\mathcal I} (M) = \{ \bth \}$, $M$ is structurally globally identifiable (SGI);
\item  the elements of  ${\mathcal I}(M)$ are denumerable,  $M$ is structurally locally identifiable (SLI); 
\item  the elements of ${\mathcal I}(M)$ are not denumerable, $M$ is structurally unidentifiable (SU).
\end{description}
\end{definition}

We note that some care is needed in the application of Definition \ref{def:ID}, as it is not appropriate in all cases.
Condition 1 ensures that the definition is not applicable to all classes of systems, including switching systems. 
Condition 2 indicates that the initial state cannot be an equilibrium point,  as otherwise response is constant for all time. Such a response cannot provide information on system dynamics. If the constant response is atypical, it does not provide an appropriate idealisation of real data. Thus, it is inappropriate to use a constant response in testing the structure for SGI. 

\begin{remark}
Instead of the test described above, one may test a structure for the property of structural local identifiability
(\cite{Structural_Villaverde_2016}).
This is able to judge a structure as either SLI, or SU. 
Discerning that a structure is SLI may be adequate in some circumstances, and the tests tend to be 
easier to apply than tests for SGI.
\end{remark}

In general, the output of system $M(\bth)$ features ``(structural) invariants'' \cite{Vajda_Structural_1981}
 (or ``observational parameters'' \cite{Jacquez_id85}) $\bph(\bth)$ which define the time course of output.  We may 
 use these to summarise the properties of 
 the whole structure.\footnote{We can conceive of invariants most directly when a structure is defined by one
 set of mathematical relations for all time. Otherwise, say for structures of switching systems, we require
a more flexible approach (\cite{MABE07, jwhyte_On_det_07}). Such structures are beyond the introductory
 intentions of this chapter.}

Thus, invariants allow us to test a structure for SGI using algebraic conditions that are addressed more easily than a functional relationship as in~\eqref{eq:ID_def}. Here we formalise this property by rewriting Definition~\ref{def:ID} in terms of invariants. This leads to a test of
a structure for SGI that is easier to apply than its predecessor.

\begin{definition} \label{def:ID_invariants}
Suppose that structure $M$ satisfies Conditions~1 and 2 of Definition~\ref{def:ID}. Then, for some arbitrary $\bth \in \Theta$, we define the set
\begin{align}
 {\mathcal I}(M,\bph) \triangleq \left\{ \boldsymbol{\theta^{'} } \in \Theta: 
 \boldsymbol{\phi(\theta^{'})} = \bph(\bth)  \right\} \equiv   {\mathcal I}(M) \; .
\label{eq:ID_def2}
\end{align}
It follows that determination of $ {\mathcal I}(M,\bph)$ allows classification of $M$ according to Definition \ref{def:ID}.
\end{definition}

Given Definition~\ref{def:ID_invariants}, we may propose a process for testing a structure for SGI.

\begin{proposition} \label{prop:ID_test}
\begin{description}
\item[]
\item[Step 1] Obtain invariants $\bph(\bth)$:
there are various approaches, but some have requirements (e.g. that the structure is generically minimal) that may be difficult to check. 
\item[Step 2] Form alternative invariants $\bph(\bth^{\prime})$ by substituting $\bth^{\prime}$ for $\bth$
in $\bph(\bth)$.
\item[Step 3] Form equations $\bph(\bth^{\prime})= \bph(\bth) $.
\item[Step 4] Solve equations.
\item[Step 5] Scrutinise solution set to make a judgement on $M$ according to Definition~\ref{def:ID_invariants}.
\end{description}
\end{proposition}

Step 1 poses a key problem : how may we obtain some suitable $\bph$? When considering an LTI structure,
the TFA is appropriate. We will now introduce the approach, proceeding to illustrate its application to an uncontrolled LTI structure in Sect.~\ref{ss:uLTI_ex}.

\subsection{The Transfer Function Approach}
Consider a compartmental LTI structure $S$ with indices $n,k \in \mathbb{N}$ and $m \in \mathbb{N}_{0}$,
having system matrices belonging to $SL \Sigma P(n,m,k)$ (recalling that $m=0$ indicates an uncontrolled structure).
Recall the idealised framework employed in the testing of a structure for SGI shown in Convention~\ref{conv:ID_framework}.
As such, we consider $S$ defined for time set $T=\bar{\mathbb{R}}_{+}$ .
Recall \eqref{eq:LT_controlled_y}, and the discussion of Sect.~\ref{sss:time_course} which guarantees that there exists
some $\lambda$ such that the Laplace transform of $\vec{y}$ has a domain of convergence.
Then, given transfer matrices $\vec{V}$ and $\vec{W}$ (as appropriate), we may extract invariants for use in 
testing $S$ for SGI.  First, we must place the transfer functions into a specific form.

\begin{definition}[Canonical form of a transfer function] 
 \label{defn:canonical_form}
Given compartmental LTI structure $S$ of $n \in \mathbb{N}$ states, suppose that associated with $S(\bth)$ 
is a transfer matrix (as in \eqref{eq:LT_controlled_y}) $\vec{Z}$, composed of unprocessed transfer functions. 
Given element $z_{i,j}(s;\bth) \in \mathbb{C}(s)$, we obtain the 
associated {\it transfer function in canonical form} by cancelling any common factors between the numerator and denominator, and rewriting to ensure 
that the denominator polynomial is monic. 
The result is an expression of the form:
\begin{gather} 
 \begin{gathered}
 z_{i,j}(s;\bth)  = 
 \frac{\omega_{i,j,r+p}(\bth) s^{p} + \cdots + \omega_{i,j,r}(\bth) }{s^{r} + \omega_{i,j,r-1}(\bth) s^{r-1}+ \cdots + 
  \omega_{i,j,0}(\bth)}, \quad \forall s \in \mathbb{C}_0 \supseteq H_{\lambda} \; ,  \\
 r \in \{ 1, \dots, n\} \; , \quad   p \in \{ 0, \dots,  r-1\} \; .
 \end{gathered} \label{eq:LT_output}
\end{gather}
The coefficients $\omega_{i,j,0}, \ldots, \omega_{i,j,r+p}$ in \eqref{eq:LT_output} contribute invariants towards $\bph(\bth)$. 
\end{definition}

\subsection{A demonstration of the testing of an uncontrolled LTI structure for SGI} \label{ss:uLTI_ex}

Recalling the general form of systems in an uncontrolled compartmental LTI structure from  \eqref{eq:LTI_ode_state} and \eqref{eq:LTI_y},
let us consider a particular example $S_{0}$, with representative system:
\begin{gather}
 \vec{\dot{x}_{0} }  (t;\bth)  = {\bf A}(\bth)\cdot {\bf x_{0}} (t;\bth) \; \quad {\bf x_{0}}(0;\bth) = {\bf x_{0_0} }(\bth) \; ,
\label{eq:S1_state} \\
  y_{0}(t;\bth) = {\bf C }(\bth) \cdot {\bf x_{0}} (t;\bth) \; , \label{eq:S1_y1}
\end{gather}
where the state vector is
${\bf x_{0}}(t;\bth) = \begin{bmatrix} x_{1} & x_{2} & x_{3} \end{bmatrix}^{\top}$, and the system matrices belong to $SL\Sigma P(3,0,1)$. These have the form:
\begin{align}
\begin{aligned}
{\bf x_{0_0}}(\bth)  = \begin{bmatrix} 0 \\ x_{20} \\ 0 \end{bmatrix}, \quad
{\bf A}(\bth) = \begin{bmatrix} -k_{01} - k_{21} & k_{12} & 0 \\
k_{21} & -k_{12}-k_{32} & k_{23} \\                        
0 & k_{32} & -k_{23} \end{bmatrix} , \quad 
{\bf C}(\bth)  = \begin{bmatrix}  1 & 0 & 0\end{bmatrix}  , 
\end{aligned}
\end{align}
and we have parameter vector 
\begin{align}
\bth &= \left( k_{01} ,  k_{12} ,  k_{21} ,  k_{23} ,  k_{32} , x_{20}   \right)^{\top} \in \mathbb{R}^{5}_{+} \; . \label{eq:theta}
\end{align}

Condition~1 of Definition~\ref{def:ID} is satisfied for linear systems. To test whether $S_{0}$ satisfies Condition~2 of Definition~\ref{def:ID}, we note that
\begin{align}
\begin{aligned}
{{\vec{\dot{x}_{0}}} }(0,\bth) &= \vec{A}(\bth) \vec{x_{0_0}}(\bth) =  \begin{bmatrix} k_{12} x_{20} \\ -(k_{12} + k_{32})x_{20} \\ k_{32}x_{20} 
					\end{bmatrix} \\
					& \neq \vec{0} \quad (\mbox{as all parameters are strictly positive}),
\end{aligned}
\end{align}
and thus the condition is satisfied for all $\bth \in \Theta$. As the conditions of Definition~\ref{def:ID}
are satisfied, we may proceed in testing $S_{0}$ for
SGI following Proposition~\ref{prop:ID_test} and Definition~\ref{def:ID_invariants}.
 
Recall that in this uncontrolled case, the Laplace transform of the output function has the form of 
\eqref{eq:LT_uncontrolled_y}. Following the notation introduced earlier, we write the transform for $y_{0}(\cdot; \bth)$ 
as  $\prescript{S_0}{}{V(s;\bth)}$, which is a scalar, and the only source of invariants for $S_{0}$. Deriving the expression (and neglecting 
the matrix indices of Definition~\ref{defn:canonical_form} for simplicity) yields
\begin{gather} 
\begin{aligned} 
 \prescript{S_0}{}{V(s;\bth)} = 
 \frac{  \phi_{4}(\bth) s +  \phi_{3}(\bth) }{s^{3} + \phi_{2}(\bth) s^{2} + \phi_{1}(\bth) s + \phi_{0}(\bth)} \; , 
 \quad \forall s \in \mathbb{C}_0, \\
\end{aligned}  \label{eq:S1_V_ratfn} \\
\intertext{where}
\begin{aligned}
\phi_{0}(\bth) &= k_{01} k_{12} k_{23} \; , \\
\phi_{1}(\bth) &= k_{01} k_{12} + k_{01} k_{23} + k_{01} k_{32} + k_{12} k_{23} + k_{21} k_{23} + k_{21} k_{32} \; , \\
\phi_{2}(\bth) &= k_{01} + k_{12} + k_{21} + k_{23} + k_{32} \; , \\
\phi_{3}(\bth) & = k_{12} k_{23} x_{20} \; , \\
\phi_{4}(\bth) & = k_{12} x_{20}  \; .
\end{aligned}
  \label{eq:S1_V}
\end{gather}

We set 
\begin{align}
\bph_{\bf 0}(\bth) \triangleq \Big( \phi_{0}(\bth),  \phi_{1}(\bth), \phi_{2}(\bth), \phi_{3}(\bth), \phi_{4}(\bth) \Big)^{\top} ,
\label{eq:S1_phi}
\end{align}
 and defining
\begin{align}
\bth^{\prime} & \triangleq \left( k_{01}^{\prime} ,  k_{12}^{\prime} ,  k_{21}^{\prime} ,  k_{23}^{\prime} ,  k_{32}^{\prime} , x_{20}^{\prime}   \right)^{\top} \in \mathbb{R}^{5}_{+} \;  \label{eq:theta_prime}
\end{align}
allows us to form the test equations 
\begin{align}
\bph_{\bf 0}(\bth^{\prime})  = \bph_{\bf 0}(\bth) \; .  \label{eq:y1_IDeqns}
\end{align}

We have six parameters, and merely five conditions. As such, we expect that $S_{0}$ is not SGI.
Solving System \eqref{eq:y1_IDeqns} for feasible $\bth^{\prime}$ yields the solution set:
\begin{align}
&  \mathcal{I}(S_{0}, \bph_{\bf 0} ) =  \notag \\
& \qquad \left\{ \bth^{\prime} \in \mathbb{R}^{5 }_{+} \left| 
  \begin{aligned}
&  \left\{  \begin{NiceMatrix}[columns-width = 1cm] \frac{ x_{20}^{\prime} k_{01} } { x_{20} } , &
\frac{ k_{12} x_{20} } { x_{20}^{\prime}  } , & 
 \Psi - \frac{ \sqrt{\Pi} } {2 x_{20}^{\prime} }   , &
  k_{23},  & 
  \frac{ \chi + \sqrt{\Pi} } {2x_{20}^{\prime} } ,  & 
   x_{20}^{\prime}  \end{NiceMatrix}  \right\}  , \\
 & \left\{  \begin{NiceMatrix}[columns-width = 1cm] \frac{ x_{20}^{\prime} k_{01} } { x_{20} } , & 
 \frac{ k_{12} x_{20} } { x_{20}^{\prime}  } , &
 \Psi + \frac{ \sqrt{\Pi} } {2x_{20}^{\prime} }, &
  k_{23},   &
  \frac{ \chi - \sqrt{\Pi} } {2x_{20}^{\prime} } ,   &
    x_{20}^{\prime}   \end{NiceMatrix} \right\}  
\end{aligned}   \right. \right\}    \, ,  \label{eq:soln_families_S1}
\end{align}  
where we interpret $x_{20}^{\prime}$ as a free parameter,
\begin{align}
 \begin{aligned}
 \Psi & \triangleq  \frac{ \phi_{1}(\bth) } {2} - \frac{ k_{01}x_{20}^{\prime} } { x_{20} } -
  \frac{ k_{12}x_{20} } { 2x_{20}^{\prime} }  , \qquad
\mbox{and setting $ \Xi  \triangleq  k_{01} + k_{21} - k_{23} $ allows us to write} \\
\chi & \triangleq ( \Xi + k_{12} +k_{32} ) x_{20}^{\prime} -k_{12}x_{20} ,\\
\Pi & \triangleq \left( \Xi^{2}  + 2 (k_{12} - k_{32})\Xi + (k_{12} + k_{32})^{2} \right) x_{20}^{\prime^{2}} 
 - 2k_{12} x_{20} ( \Xi - k_{12} + k_{32}) x_{20}^{\prime}  + k_{12}^{2} x_{20}^{2} \, .
\end{aligned} \label{eq:repeating_terms}
\end{align}

By substituting $x_{20}^{\prime}= x_{20}$ into either of the solution families given in \eqref{eq:soln_families_S1}
we see that the trivial solution $\bth^{\prime} = \bth$ is also valid, as we would expect. We note that the parameter $k_{23}$ is SGI.

Even though structure $S_{0}$ contains relatively simple models, \eqref{eq:soln_families_S1} with \eqref{eq:repeating_terms} show
that the solutions for $\bth^{\prime}$ in terms of $\bth$  are somewhat complicated, and not particularly easy to categorise. 
However, we see in  \eqref{eq:soln_families_S1} that there are two distinct families of solutions. As $x_{20}^{\prime}$ is free in each, 
there are uncountably infinitely-many feasible vectors $\bth^{\prime}$ that reproduce
the structure's output for a nominated $\bth$. As such, we judge $S_{0}$ as SU.

\section{Testing controlled structures for structural global identifiability} \label{s:controlled_LTI_ID}

In considering the properties of a controlled state-space structure, we must account for the effects of inputs.
 Returning to the
testing overview outlined in Proposition~\ref{prop:ID_test}, it is appropriate to precede Step 1 with a new
step:
\begin{description}
\item[Step 0] Specify the set of inputs which may be applied to the structure.
\end{description}

It is also appropriate for us to adapt the definitions that suit uncontrolled structures for this setting.

\begin{definition} \label{def:controlled_ID}
Suppose we have controlled state-space model structure $M$ having parameter set $\Theta$ and set of input functions $\cal{U}$,
and time set $T \subseteq [0, \infty)$. For some unspecified parameter vector and input, $\bth \in \Theta$ and $\vec{u} \in \cal{U}$ respectively, we 
 illustrate $M$ with representative model $M(\bth)$ (say, as in \eqref{eq:general_controlled}), having 
  state function $\vec{x}(\cdot, \vec{u}; \bth) \in \mathbb{R}^{n}$ and
 output function  ${\bf y}(\cdot, \vec{u}; \bth) \in \mathbb{R}^{k}$. 

 Suppose that for each $\vec{u} \in \mathcal{U}$ systems in $M$ satisfy conditions:
\begin{enumerate}
\item Functions $\vec{f}(\vec{x}, \vec{u}, \cdot; \bth)$ and $\vec{g}(\vec{x}, \vec{u}, \cdot; \bth)$  are real and analytic for every $\bth \in \Theta$
on $\mathcal{S}$ (a connected open subset of $\mathbb{R}^{n}$  such that $\vec{x}(t, \vec{u}; \bth) \in \mathcal{S}$ for every
$t \in [0, \tau]$, $\tau >0$).
\item For $t$ belonging to (at least) some subinterval of $[0, \tau]$, $\vec{f}(\vec{x}, \vec{u}, t ;  \bth) \neq \vec{0}$ for almost all $\bth \in \Theta$. 
\end{enumerate}

 Given finite time $\tau >0$, we define
\begin{align}
\displaystyle {\mathcal I}(M, \mathcal{U}) \triangleq \left\{ \boldsymbol{\theta^{'}} \in \Theta: 
{\bf y}(t, \vec{u}; \boldsymbol{\theta^{'}}) = {\bf y}(t , \vec{u}; \bth)  \quad \forall t \in [0,\tau], \ \forall \vec{u} \in \mathcal{U}  \right\} . \label{eq:ID_def_controlled}
\end{align}
If, for almost all $\bth \in \Theta$: 
\begin{description}
\item $   {\mathcal I} (M, \mathcal{U}) = \{ \bth \}$: $M$ is structurally globally identifiable for input set $\mathcal{U}$ ($\cal{U}$-SGI);
\item the elements of  ${\mathcal I}(M, \mathcal{U})$ are denumerable:  $M$ is structurally locally identifiable for input set 
$\mathcal{U}$ ($\cal{U}$-SLI); 
\item the elements of ${\mathcal I}(M, \mathcal{U})$ are not denumerable: $M$ is structurally unidentifiable for 
input set $\cal{U}$ ($\cal{U}$-SU).
\end{description}
\end{definition}

\begin{remark}
Conditions 1 and 2 of Definition~\ref{def:controlled_ID} play similar roles to the corresponding conditions of Definition~\ref{def:ID}.
Condition 1 excludes from consideration structures subject to discontinuities in the state or output functions, for which we cannot readily define invariants. Condition 2 relates to conditions which allow us to elicit informative input from a system in $M$.
This loosens the condition of the uncontrolled case, where a system at equilibrium at $t=0$ remains there.
The controlled case is different; a system at an equilibrium state may be displaced by the action of an input. However,
this alone does not guarantee that the output of a controlled system is informative for any input in $\mathcal{U}$. As such, Condition 2 seeks to preclude the case where the system's state is largely constant, possibly changing only at isolated points on $[0,\tau]$. By doing so, we expect to obtain useful (non-degenerate) output, and possibly, invariants subsequently, depending on the nature of $\mathcal{U}$.

Should Conditions 1 and 2 not hold for any $\vec{u} \in \mathcal{U}$, it is appropriate to remove these
from the input set.
\end{remark}

Suppose $M$ satisfies Conditions 1 and 2 of Definition~\ref{def:controlled_ID}, and we may observe $M$'s outputs for $\mathcal{U}$ containing a sufficiently broad range of inputs (e.g. the set of piecewise continuous functions defined on $T$, \cite{Vajda_Structural_1981}). Then,
within our idealised testing framework (Convention~\ref{conv:ID_framework}) we can access the 
structure's invariants, say $\bph$. In such a case, 
rather than making a judgement on $M$ using Definition~\ref{def:controlled_ID},  we may use $\bph$ with the more convenient Definition~\ref{def:ID_invariants}. 

Let us turn our attention to the application of Definition~\ref{def:controlled_ID} when $M$ is a controlled compartmental LTI structure. By physical reasoning ($\vec{x}$ is real and does not exhibit jumps, and these properties are transferred to $\vec{y}$) we expect that Condition 1 is satisfied. Checking Condition 2 may not be trivial in general, and so it may be easier to verify an alternative condition, even if this is stricter than necessary. For example, 
if we were  to show that $\vec{\dot{x}}(t;\bth) \neq \vec{0}$
for almost all $\bth \in \Theta$ and any $t \in [0, \tau]$ for finite $\tau$, then Condition 2 is satisfied.

In practice, conditions such as those of Definition~\ref{def:controlled_ID} do not typically feature in
discussions of the testing of controlled LTI structures for SGI. This is likely due to the expectation that 
one can access a structure's invariants if the input set meets only modest requirements: that $\mathcal{U}$ is sufficiently diverse, and that the Laplace transform of any input in $\mathcal{U}$ exists. Satisfying these conditions allows
us to derive transfer matrices  ${\bf W} $  and ${\bf V }$ as in \eqref{eq:LT_controlled_y}, place transfer functions contained therein in canonical form (recall Definition~\ref{defn:canonical_form}),  and obtain $\bph$ from their coefficients.

In various situations, for practical or ethical reasons, one is limited in the nature and number of inputs that one can apply to some physical system. In such a case, it is not appropriate to assume that we may access $\bph$ from $M$.  As such, the testing framework 
seen in Definition~\ref{def:ID_invariants} is an inappropriate idealisation. However, we may consider the result of such a test
as a ``best case scenario''---we would not expect to obtain a more favourable result from a limited set of inputs. 
As such, if a test using $\bph$ shows that $M$ is SU, we can be 
almost certain that PI applied to the output from our physical system resulting from a limited set of inputs will not obtain unique parameter estimates.
Inconveniently, when the test classifies $M$ as SGI or SLI, we cannot necessarily ascertain whether this
judgement will also apply when we know that limited inputs are available. As such, it is appropriate to 
return to Definition~\ref{def:controlled_ID} and consider a test for generic uniqueness of parameter vectors that takes into account the set of available inputs, and which does not require invariants.

Some authors have noted situations where---unlike in the testing of a structure for SGI based on
invariants---we may not consider inputs as being applied sequentially to yield separate output time courses.
For example, in considering LTI compartmental structures, Godfrey~\cite[Page~95]{Godfrey_book} cautioned:
\begin{quotation}
However, when more than one input is applied simultaneously, identifiability may depend on the shape of the two
inputs, and it is then essential to examine the form of the observations $\vec{Y}(s)$ [the Laplace transform of $\vec{y}$]
rather than individual transfer functions. 
\end{quotation}
In noting the importance of the available set of inputs,
Jacquez and Grief~\cite[Page~201]{Jacquez_id85} sought to distinguish ``system identifiability'' (which we understand as SGI) 
from ``model identifiability'' which depends on some particular inputs (as we have allowed for in Definition~\ref{def:controlled_ID}). The authors noted the confusion caused by failing to distinguish between these different properties. To the best of our knowledge, the literature does not have consistent terminology to distinguish these concepts, which may be a consequence of 
how infrequently it is explicitly considered.

We will seek to reuse the TFA machinery in considering what parameter information we may glean from the idealised output of a compartmental LTI structure subject to a single input. 
Let us consider such a structure $S$ having system matrices in  $SL\Sigma P(n,m,k)$. 
Suppose that we can observe idealised output for a single input $\vec{u}$, that is $\mathcal{U} = \{ \vec{u} \}$,
and that $\mathcal{L} \{ \vec{u} \}(s)$ exists.
Then, we may obtain parameter information for testing $S$ for SGI given $\vec{u}$ from
\begin{align}
\mathcal{L} \{ y \}(s;\bth) = {\bf C}(\bth) \big( s {\bf I} - {\bf A}(\bth) \big)^{-1} 
\big(  {\bf x_{0}(\bth)} + {\bf B}(\bth) \mathcal{L}\{ {\bf u} \}(s)  \big) \; .
\end{align}

In order to demonstrate the difference between the testing of a controlled structure when invariants are and are not obtainable, we
shall consider an example structure for which different input sets are available. Recall the SISO structure $S_{1}$ from 
Sect.~\ref{sec:1}. Following definitions from Sect.~\ref{s:prelims}, we rewrite the representative system in state-space form as
\begin{gather}
\dot{ {\bf x_{1} } } (t;\bth)  = {\bf A}(\bth)\cdot {\bf x_{1}} (t;\bth) + {\bf B}(\bth)u(t) \; , \quad 
{\bf x_{1}}(0;\bth) = {\bf x_{1_0} }(\bth) \; ,   \label{eq:S2_state} \\
  y_{1}(t;\bth) = {\bf C }(\bth) \cdot {\bf x_{1}} (t;\bth) \; , \label{eq:S2_y}
\end{gather}
where the state vector is ${\bf x_{1}}(t;\bth) = \begin{bmatrix} x_{1} & x_{2} & x_{3} \end{bmatrix}^{\top}$, and system matrices belong to $SL \Sigma P(3,1,1)$. Specifically we have
\begin{align}
\begin{gathered}
{\bf x_{1}}(0;\bth) = \begin{bmatrix} 0 \\ x_{20} \\ 0 \end{bmatrix}, \quad 
{\bf A}(\bth) = \begin{bmatrix} -k_{01} - k_{21} & k_{12} & 0 \\
k_{21} & -k_{12}-k_{32} & k_{23} \\                        
0 & k_{32} & -k_{23} \end{bmatrix} , \\
{\bf B}(\bth)  = \begin{bmatrix}  0 \\ 0 \\ 1 \end{bmatrix} , \quad 
{\bf C}(\bth)  = \begin{bmatrix}  1 & 0 & 0\end{bmatrix}  .
\end{gathered}
\end{align}
Recalling  \eqref{eq:V} and \eqref{eq:W}, the transfer matrices here are scalars, which henceforth we denote by
 $\prescript{S_{1}}{}{W}$ and $\prescript{S_{1}}{}{V}$.
We note that by neglecting $\vec{B}$ we obtain the uncontrolled LTI structure $S_{0}$ (recall \eqref{eq:S1_state} and \eqref{eq:S1_y1}). Structure $S_{1}$ has the same parameter vector as $S_{0}$, shown in \eqref{eq:theta}.

Below we proceed to test $S_{1}$ for SGI under the assumption that we can obtain its invariants.

\subsection{A demonstration of the testing of a controlled LTI structure for SGI when invariants are accessible from outputs}
\label{ss:S1_invariant_test}
Let us assume that we have the idealised outputs of $S_{1}$ for a sufficiently large input set $\mathcal{U}$ such that
we can obtain $\prescript{S_{1}}{}{W}$ and $\prescript{S_{1}}{}{V}$. By converting 
each of these rational functions into the canonical form,  we may obtain each coefficient of $s$.
The collection of these specifies a vector of invariants. We shall recall the steps of Proposition~\ref{prop:ID_test} in testing 
$S_{1}$ for SGI.

Towards Step 1, those invariants relating to the response due to the initial conditions reside in 
$ \prescript{S_{1}}{} {V} \equiv \prescript{S_{0}} {} {V}$. We collected these invariants in \eqref{eq:S1_phi}. 

The behaviour of $S_{1}$ differs from that of $S_{0}$ due to the invariants relating to inputs, held in $ \prescript{S_{1}}{} {W}$.
Following \eqref{eq:W}, we see that  $\prescript{S_{1}} {} {W} \triangleq {\bf C}(s{\bf I}_{3}  - {\bf A})^{-1}{\bf B}$, from
which we obtain the transfer function in canonical form:
\begin{align}
\prescript{S_{1}}{} {W}(\bth) = \frac{ \omega_{0}(\bth) } {s^{3} + \phi_{2}(\bth)s^{2} + \phi_{1}(\bth)s + \phi_{0}(\bth) } \; ,
\label{eq:S2_W}
\end{align} 
where the denominator invariants repeat the corresponding coefficients in 
$\mathcal{L}\{ y_{0} \}(s; \bth)$ (recall \eqref{eq:S1_V}), and
\begin{align}
\omega_{0}(\bth) & = k_{12} k_{23} \; .  \label{eq:omega0}
\end{align}
Thus, only $\omega_{0}(\bth)$ provides an invariant that is novel compared to those from $\prescript{S_{0}} { }{ V(\bth)}$.

Drawing on \eqref{eq:S1_phi} and \eqref{eq:omega0}, we complete Step 1 by forming the vector of distinct invariants associated with $S_{1}$: 
\begin{align}
\bph_{\bf 1}(\bth) \triangleq ( \underbrace{\phi_{0}(\bth), \phi_{1}(\bth), \phi_{2}(\bth)}_{ \substack{ \text{common to} \\ 
 \prescript{S_{1}}{} {V}(\bth) ,  \;  \prescript{S_{1}}{} {W}(\bth) \\ \text{denominators}}},  \underbrace{ \phi_{3}(\bth),
 \phi_{4}(\bth)}_{\substack{ \text{from numerator} \\ \text{of } \prescript{S_{1}}{} {V}(\bth)} }, 
\underbrace{\omega_{0}(\bth)}_{\substack{ \text{from numerator} \\ \text{of } \prescript{S_{1}}{} {W}(\bth)} } )^{\top} \; .  \label{eq:S1_invariants}
\end{align}
Following Step 2 we use $ \bph_{\bf 1}(\bth)$ from \eqref{eq:S1_invariants} to form the invariants dependent on our alternative parameter $\bth^{\prime}$, (as in \eqref{eq:theta_prime}), $\bph_{\bf 1}(\bth^{\prime})$. Step 3 directs us to form the test
equations $\bph_{\bf 1}(\bth^{\prime}) = \bph_{\bf 1}(\bth)$. 
Upon solving for feasible $\bth^{\prime}$  we obtain 
\begin{align}
& \mathcal{I}(S_{1}, \bph_{\bf 1})  =  \notag \\
& \qquad \left\{ \bth^{\prime} \in \mathbb{R}_{+}^{5} \left| 
\begin{aligned}
& \left\{ \begin{NiceMatrix}[columns-width = 0.65cm] \displaystyle k_{01} , &
  k_{12} ,  & \rlap{$k_{21},$}   \hphantom{-k_{01} + k_{23} + k_{32},} &  k_{23}, &
   \rlap{$k_{32},$}   \hphantom{k_{01} + k_{21} - k_{23},}  & x_{20} \end{NiceMatrix}   \right\} ,   \\ 
& \left\{  \begin{NiceMatrix}[columns-width = 0.65cm]  k_{01}  , &
k_{12} , & 
-k_{01} + k_{23} + k_{32}   , &
  k_{23}    , & 
    k_{01} + k_{21} - k_{23}, & 
   x_{20} \end{NiceMatrix}  \right\} \\
\end{aligned}  \right. \right\} . 
\label{eq:all_inputs_solns}
\end{align}

Equation \eqref{eq:all_inputs_solns} shows that we can obtain unique estimates for $k_{01}^{\prime}$, 
$k_{12}^{\prime}$, $k_{23}^{\prime}$, and $x_{20}^{\prime}$ (i.e. the corresponding true values 
in $\bth$) for any $\bth \in \mathbb{R}_{+}^{6}$. However, for each of $k_{21}^{\prime}$ and 
$k_{32}^{\prime}$ we see there are two distinct solutions whenever $-k_{01} + k_{23} + k_{32} >0$ and
$k_{01} + k_{21} - k_{23} >0$. That is, the structure is SLI. 

Inspection of the second solution family in \eqref{eq:all_inputs_solns} reveals 
$k_{21}^{\prime} + k_{32}^{\prime} =  k_{21} +  k_{32}$. This may hint that a reparameterisation of $S_{1}$
so as to replace occurrences of $k_{21} +  k_{32}$ (which may occur in combination with other parameters) with appropriate new parameters would produce a new structure which is SGI. Whilst there are techniques for
generating alternative structures that produce the same output (e.g. \cite{Walter_Unidentifiable_1981}), in general,
 finding a suitable reparameterisation amongst these is not a trivial undertaking. Given this, we may have to find some means of managing an SU structure. For example, we may determine bounds on the values of parameters by testing for ``interval identifiability''. If the bounds are sufficiently narrow, we may tolerate an SU structure (see \cite{Godfrey_Identifiability_1985} for examples).

We shall now consider $S_{1}$ in the more restrictive setting where our idealised output results from 
the application of one specific input.

\subsection{A demonstration of the testing of a controlled LTI structure for SGI when invariants are not accessible from outputs}

Suppose that we can only observe the idealised output of $S_{1}$ for the single input $u = \delta(t-0)$---the impulsive input at time zero.
Noting that $\mathcal{L}\{ \delta(t-0) \}(s) = 1$, and recalling \eqref{eq:LT_controlled_y}, we may write 
\begin{equation}
\mathcal{L} \{ y_{2}(\cdot,\bth) \}(s) =   \prescript{S_1}{} { V  }(s;\bth) + \prescript{S_1}{}{W }(s;\bth) ,
\label{eq:LT_y2}
\end{equation} 
where the terms on the right-hand side are given by \eqref{eq:S1_V_ratfn} (recalling
 $ \prescript{S_1}{}{V(s;\bth)} \equiv \prescript{S_0}{}{V(s;\bth)}$) and  \eqref{eq:S2_W}, 
respectively. 

The sum of the two transfer functions on the right-hand side of \eqref{eq:LT_y2} is also a rational function in $s$, and hence is analogous to a transfer function. As such, it is convenient to process this in a manner similar to that shown in 
Sect.~\ref{ss:S1_invariant_test}. Thus, ensuring that the right-hand side of \eqref{eq:LT_y2} is in the canonical form, and simplifying, yields an expression (which is similar to the canonical form of  $\mathcal{L} \{y_{0}(\cdot,\bth) \}(s)$, recall \eqref{eq:S1_V_ratfn}):
\begin{align}
\mathcal{L} \{ y_{2} \}(s; \bth) & = 
\frac{  \phi_{4}(\bth) s +  \beta(\bth)  }{s^{3} + \phi_{2}(\bth) s^{2} + \phi_{1}(\bth) s + \phi_{0}(\bth)}, 
 \quad \forall s \in \mathbb{C}_0 ,   \label{eq:LT_controlled_output} \\
\intertext{where, recalling \eqref{eq:S1_V} and \eqref{eq:omega0},}
\beta(\bth) & \triangleq  \phi_{3}(\bth) + \omega_{0}(\bth) =  k_{12}k_{23}  (x_{20} + 1) \, . \notag
\end{align}

\begin{remark}
Given the input $u = \delta(t-0)$ and that $S_{1}$ is an open system, mass present in the system due to the input and initial conditions is lost to the environment over time. As $t \rightarrow \infty$, the system approaches its steady state $\vec{x}^{*}= \vec{0}$. 
We note that \eqref{eq:LT_controlled_output} is the Laplace transform of an output function that is a sum of exponentials in $t$ (recall Sect.~\ref{sss:time_course}) as a result of being a linear combination of the individual state variables.
As all $\bth$ are positive, all invariants in \eqref{eq:LT_controlled_output} are also positive. As such, we
see that $\vec{y}$ is not constant. We infer that the state function $\vec{x}$ is time-varying, and that it leads
to an informative output function. Thus, $S_{1}$ for this $u$ satisfies Condition 2 of Definition~\ref{def:controlled_ID}.
\end{remark}

We note that $\mathcal{L} \{ y_{2} \}(s;\bth)$ and $\mathcal{L} \{ y_{0} \}(s;\bth)$ differ only in
the constant term of their numerators. The coefficients in 
\eqref{eq:LT_controlled_output} play a similar role to invariants as they determine the output. As a further
conceptual and notational convenience, we write
\begin{equation*}
\bph_{\bf 2}(\bth) \triangleq ( \phi_{0}(\bth), \phi_{1}(\bth), \phi_{2}(\bth), \beta(\bth) , \phi_{4}(\bth))^{\top}.
\end{equation*}
Following Steps 2 and 3 of Proposition~\ref{prop:ID_test} leads to a system of test equations 
$\bph_{\bf 2}(\bth^{\prime}) = \bph_{\bf 2} (\bth)$, containing four of the five equations used in testing $S_{0}$ for SGI.

Let us consider the difference between the systems of equations which follow from $\bph_{0}$ and $\bph_{2}$. 
The analysis of $S_{0}$ produces a novel equation involving $\phi_{3}$. In analysing $S_{1}$ output due to a single input here,
the novel equation is due to $\beta(\bth)$. This allows $k_{12}$, $k_{23}$, and $x_{20}$ more freedom than that permitted by 
the  $\phi_{3}$ equation. Thus, solving $\bph_{\bf 2}(\bth^{\prime}) = \bph_{\bf 2} (\bth)$ yields an even more complicated solution set than that seen for $S_{0}$ in \eqref{eq:soln_families_S1} and \eqref{eq:repeating_terms}. As a kindness to the reader, 
we shall not present the solution sets here. However, classification of the structure is straightforward as 
$ \bph_{\bf 2} (\bth)$ provides five equations, yet we have six parameters.
Thus, when the input set is $\mathcal{U} = \{ \delta(t-0) \} $, we classify $S_{1}$ as $\mathcal{U}$-SU.

This is a less-favourable result than the classification of $S_{1}$ as SLI (recall the the assumption that outputs are
available for a broad enough range of inputs) as demonstrated in \eqref{eq:all_inputs_solns}. This result reinforces the claim that, when intending to test a structure for SGI, it is appropriate to specify the inputs which will be applied to physical system. Thence, we may judge whether or not the associated idealised output allows determination of invariants, and
use this knowledge in choosing an appropriate testing method.

\section{Concluding remarks} \label{s:conclusions}
This overview has aimed to highlight the benefits of testing model structures for the property of structural global 
identifiability (SGI). 
Moreover, by assembling crucial definitions, drawing important distinctions, and providing test examples, we have sought to
illuminate some important concepts in the field of identifiability analysis. We hope that this will encourage and assist 
interrogation of proposed structures so as to recognise those that are not SGI. This will allow
researchers to anticipate the frustrations almost certain to accompany the use of a non-SGI structure
(especially, an unidentifiable one) in modelling and parameter estimation. 

Progress in the field of identifiability analysis is ongoing through the development of new methods of testing structures for SGI or SLI, and refinements to their implementation. However, certain practical matters are yet to receive widespread consideration. 
We conclude with brief comments on a selection of these.

\paragraph{Competition---or collaboration---between testing methods?}
Over a period of time, the literature has reported that one cannot generally anticipate which method will be easiest to apply to a given case, (e.g. \cite[Page~96]{Godfrey_book}), or that testing methods may suit some problems more than others (e.g. \cite{Chis_Structural_2011}). Consequently, when considering 
software implementations of testing methods, we may not be able to anticipate which method will produce a result in the shortest time, or at all. This uncertainty has prompted various comparisons aimed at evaluating the utility of alternative methods for testing structures for SGI. 

 One may wonder if a competitive treatment of methods is a limiting one. 
That is, might there be benefits in combining methods so as to draw upon their strengths? For example, in considering controlled compartmental LTI structures, the TFA provides a means of ascertaining whether or not a structure
is generically minimal. If the conclusion is positive, we may then change our approach and apply a suitable testing method that uses 
a type of invariant expected to be simpler than those used in the TFA. For example, we may choose Markov and initial parameters
as invariants, expecting these polynomials in the parameters to have a lower degree than those seen in transfer function coefficients.  Given such simpler invariants, the resulting test equations will have a reduced algebraic complexity. We could reasonably 
expect to solve these more quickly than equations obtained from the TFA.

\paragraph{Reproducibility of analysis}
There is a growing concern over the reproducibility of studies in computational biology (\cite{Laubenbacher_BMB_Editorial_2018}).
 We expect a greater awareness of identifiability analysis to encourage the asking of questions that will contribute to a rigorous and defensible modelling practice. Beyond this, we may also ponder how to promote reproducibility through the processes by which identifiability analysis is undertaken.

For all but the simplest cases, testing a structure for SGI requires the use of a computer algebra system (CAS). Often this is a commercial product, such as Maple\texttrademark, Mathematica, or MATLAB.
However, as for all complex computer code, one cannot necessarily guarantee that results produced by a CAS will be correct
in all situations (see, for example, \cite{Armando_reconstruction_2005} noting a limitation of certain versions of Maple\texttrademark). 
As such, it is good practice for us to check that results obtained from one CAS agree with those
from another. 

Performing such a comparison might not be straightforward.
Recall that the classical approach to testing a structure for SGI requires the solution of a system of algebraic equations.
If two CASs employ differing methods in solving a given system,
the solution sets may appear quite dissimilar, even if they are, in fact, the same. This complicates the 
task of determining whether or not the solution sets are equivalent. 

We may be able to make choices that can reduce the complexity of the comparison problem. One approach is
to seek to direct the output of CASs by specifying similar options in their commands where this is possible. For example, the ``solve'' command in Maple{\texttrademark} allows the user to specify various options, including some relating to how any solutions are displayed. Another Maple{\texttrademark} option allows some variables to be specified as ``functions of a free variable''.
We may be encouraged to use this given the form of solutions obtained from another CAS which we would like to emulate.

The seeming dissimilarity of solutions may be due to features of CAS solution algorithms that we cannot directly control. As such, we may seek to manage these by further scrutinising our equations (or more fundamentally our
invariants $\bph(\bth)$), before we attempt to solve them. 

Suppose that each (multivariate polynomial) element of $\bph(\bth)$ is some combination of simpler polynomials in  parameters $\bth$. We may determine these new polynomials by calculating a Gröbner basis for $\bph(\bth)$.\footnote{We may consider a Gröbner basis for a list of polynomials as analogous to the reduced row-echelon form of a system of linear equations.} 
This requires an ``ordering'' of parameters, which determines how terms are arranged within
a polynomial, and how monomials are arranged within terms.\footnote{For example, the polynomial
$x^{2}y + 2xy^{3} -4x +y $ employs ``pure lexicographical ordering'' with ``$x>y$''---terms are arranged by decreasing degree of monomials in $x$, and within each term any monomial in $x$ appears before one in $y$.
Changing the ordering to ``$y >x$'' yields an alternative form: $ 2y^{3}x + yx^{2} + y -4x$.}
We may obtain differing bases depending on the chosen ordering.

 In certain solution methods (such as ``nonlinsolve'' in version 1.4 of Python package SymPy) a CAS may (effectively) calculate a Gröbner basis for invariants, choosing an ordering without user input. In such cases, 
should different CASs employ differing orderings, the solutions of test equations may appear quite different. 
As such, it may be useful for the user to obtain a Gröbner basis for a specified ordering, and use this in formulating
test equations for each CAS. 

We shall illustrate the importance of the choice of ordering by returning to our example structure $S_{1}$. In Maple 2019 (version 1) we used the ``Basis'' command (from the ``Groebner'' package) to compute Gröbner bases for $\bph_{1}(\bth)$ under different orderings. We varied the ordering of parameters, as specified by the ``plex()'' option (pure lexicographical ordering). The ordering indicates a decreasing preference for eliminating parameters from our input polynomials (here, our invariants) as we proceed from the start of the list, with the aim of forming a ``triangular'' system in $\bth$. As such, those parameters occurring earlier in the list are more likely to be eliminated than those occurring later. 

Using the ordering $k_{21} > k_{32} > k_{01} > k_{12} > k_{23} > x_{20}$ yields the Gröbner basis:
\begin{align}
\vec{b}_{1}(\bth) \triangleq \begin{bmatrix} k_{{12}}x_{{20}}, \\ k_{{12}}k_{{23}}, \\
-k_{{01}}k_{{12}}+k_{{12}}k_{{32}}+k_{23}^{2}+2k_{{23}}k_{{32}}+k_{32}^{2} , \\
k_{{01}}+k_{{12}}+k_{{21}}+k_{{23}}+k_{{32}} \end{bmatrix} \, . \label{eq:G1}
\end{align}
Alternatively, with the ordering $k_{23} > k_{32} > x_{20} > k_{21} > k_{12} > k_{01}$, Maple{\texttrademark} produces the Gröbner basis:
 \begin{align}
\vec{b}_{2}(\bth) \triangleq \begin{bmatrix}
k_{01}^{2}+2k_{{01}}k_{{21}}+k_{{21}}k_{{12}}+k_{21}^{2} \\
k_{{12}}x_{{20}} \\
k_{{01}}k_{{12}}+k_{12}^{2}+k_{{21}}k_{{12}}+k_{{12}}k_{{32}} \\
k_{{01}}+k_{{12}}+k_{{21}}+k_{{23}}+k_{{32}} \, .
\end{bmatrix}  \label{eq:G2}
\end{align}
The Gröbner bases $\vec{b_{1}}(\bth)$ and $\vec{b_{2}}(\bth)$ are not identical, having only two components (the first and fourth components of $\vec{b_{1}}(\bth)$) in common. (We also note that although \eqref{eq:S1_invariants} shows 
$\bph_{1}(\bth)$ as comprised of six invariants, \eqref{eq:G1} (or \eqref{eq:G2}) shows that in the testing of $S_{1}$ for SGI,
$\bth$ is subject to only four independent conditions.)

Suppose now that---in a similar manner as we did for $\bph_{1}(\bth)$---we use $\vec{b}_{1}(\bth)$ and $\vec{b}_{2}(\bth)$ in turn to define two distinct systems of four SGI test equations. The associated solution sets for $\bth^{\prime}$, $\mathcal{I}(S_{1}, \vec{b_{1}})$ 
and $\mathcal{I}(S_{1}, \vec{b_{2}})$ respectively, determined by Maple{\texttrademark} appear to be quite different.
For example, $\mathcal{I}(S_{1}, \vec{b_{1}})$ shows $k_{12}^{\prime}$ and $k_{32}^{\prime}$ as free parameters,
whereas $\mathcal{I}(S_{1}, \vec{b_{2}})$ has $k_{01}^{\prime}$ and $k_{21}^{\prime}$ free.
This result suggests that using a Gröbner basis of our invariants to define SGI test conditions may remove one cause 
of unwanted variation between results obtained by different CASs.

When faced with (potential or actual) disparities between CAS results, access to the source code may illuminate the cause of 
the divergence, and contribute to its resolution. However, certain CAS do not permit such access to the source. In light of this, we are currently developing open-source code using the programming language Python, making particular use of the SymPy (symbolic algebra) package. By implementing this in the Jupyter notebook environment, we intend to develop implementations of testing algorithms (as we have for the TFA approach) that are readily accessible to the scientific community, and permit user customisation.

\section*{Acknowledgements}
The author is grateful to the organisers of the programme ``Influencing public health policy with data-informed mathematical models of infectious diseases'' at MATRIX (Creswick, Victoria, July  1-12 2019) for the invitation to present, and exposure to aspects of infectious disease modelling. Appreciation goes also to the organisers of the programme ``Identifiability problems in systems biology" held at the American Institute of Mathematics, San Jose, California (August 19-23 2019) and its participants, for useful discussions on contemporary problems.

  \bibliography{Whyte\_ID\_PostRev\_arXiv.bbl}

\end{document}